# A model for asymmetric giant magnetoimpedance in field-annealed amorphous ribbons

N. A. Buznikov,[a)] CheolGi Kim,[b)] Chong-Oh Kim, and Seok-Soo Yoon[c)]

*Research Center for Advanced Magnetic Materials, Chungnam National University, Daejeon 305-764, Korea*

**Abstract**

A phenomenological model for the asymmetric giant magnetoimpedance (GMI) in field-annealed amorphous ribbons is developed. The effect of a surface crystalline layer on the GMI response is described in terms of an effective bias field appearing due to a coupling between the crystalline layer and amorphous phase. It is shown that the presence of the bias field changes drastically the GMI profile. At low frequencies, the domain-walls motion leads to a steplike change in the GMI response. At high frequencies, the domain-walls motion is damped, and the GMI profile exhibits asymmetric two-peak behavior. The calculated dependences are shown to be in a qualitative agreement with results of experimental studies of the asymmetric GMI in field-annealed Co-based amorphous ribbons.

---

[a)] Permanent address: Institute for Theoretical and Applied Electrodynamics, Russian Academy of Sciences, Moscow, Russia.

[b)] Author to whom correspondence should be addressed; electronic mail: cgkim@cnu.ac.kr

[c)] Permanent address: Department of Physics, Andong National University, Andong, Korea.

Much attention has been paid recently to the asymmetric giant magnetoimpedance (GMI) effect, which is promising for the development of weak magnetic-field sensors. The asymmetric GMI has been observed first for twisted Co-based amorphous wires with dc bias current superimposed on the driving current.[1] Another method of producing the asymmetric GMI profile consists of applying an axial ac bias field.[2] A very large asymmetric GMI effect has been observed also in Co-based amorphous ribbons annealed in air in the presence of a weak magnetic field.[3–6] The asymmetry of the GMI profile has been ascribed to a hard magnetic phase, which appears due to the surface crystallization of an amorphous ribbon.[4] The coupling between crystalline and amorphous phases produces an effective bias field that is responsible for the asymmetric GMI in field-annealed ribbons.[3,5] At sufficiently low frequencies, the GMI profile exhibits a drastic steplike change in the impedance near zero field.[3,5,6] Since the behavior is similar to the magnetoresistance of spin valves, this phenomenon has been referred to as "GMI valve." At high frequencies, the GMI profile shows asymmetric two-peak behavior, with the peak at a positive field being higher than the peak at a negative field.[6] An attempt to explain the asymmetric two-peak behavior of the GMI in field-annealed ribbons has been reported in the framework of the quasistatic approach[7] and by using a model based on a solution of linearized Maxwell equations and the Landau–Lifshitz equation.[8] However, a theoretical explanation of the GMI valve phenomenon is still missing.

In this Letter, we present a phenomenological model to describe the field and frequency dependences of the asymmetric GMI in field-annealed amorphous ribbons. Both the domain-wall motion and magnetization rotation contributions to the transverse permeability are taken into account. The coupling between the surface crystalline layer and the amorphous phase is considered in terms of an effective bias field. The model proposed allows one to explain the main features of the asymmetric GMI effect, GMI valve, and asymmetric two-peak behavior, observed in field-annealed Co-based amorphous ribbons.

It is well known that the relationship between the impedance and permeability of a conductor can be described in terms of the classical skin effect. Under some simplifying assumptions, the impedance $Z$ of an amorphous ribbon can be presented in the form[9,10]

$$Z = R_{dc}[(1-i)d/2\delta] \times \coth[(1-i)d/2\delta], \qquad (1)$$

where $R_{dc}=l/\sigma dw$ is the dc ribbon resistance; $l$, $w$, $d$ and $\sigma$ are the ribbon length, width, thickness, and conductivity, respectively; $\delta=c/(2\pi\sigma\omega\mu)^{1/2}$ is the skin depth; $c$ is the velocity of light; $\omega$ is the angular current frequency; and $\mu$ is the transverse permeability. The frequency and field dependences of the transverse permeability control the GMI response of the sample. In real amorphous samples, the transverse permeability depends on many factors, such as the domain configuration, anisotropy axes distribution, the mode of the magnetization, and so on. The effect of these factors is very complex, making accurate modeling for real materials very difficult.

In the model, we assume a simplified domain structure of an amorphous ribbon, which consists of two different types of domains. Figure 1 shows schematically the geometry of the problem and the coordinate system used for the analysis. It is assumed that the ribbon has the uniaxial anisotropy with the anisotropy field $H_a$, and the anisotropy axis makes the constant angle $\psi$ with the longitudinal direction. The angle of the domain walls coincides with the anisotropy axis angle. The field annealing induces the unidirectional anisotropy in the surface crystalline layer.[4] Due to the magnetostatic or magnetoelastic coupling between amorphous and crystalline phases, an effective bias field $H_b$ appears in the amorphous region. The bias field is in the opposite direction to the unidirectional anisotropy in the surface layer.[5,6] In real ribbons, the bias field varies over the ribbon thickness. However, we consider for simplicity that the value of $H_b$ and the angle of the bias field $\varphi$ are constant over the ribbon thickness.

The equilibrium angles of the magnetization vectors in the domains, $\theta_1$ and $\theta_2$, and the equilibrium domain-wall displacement $z_0$, can be found by minimizing the free energy.[11,12] The free energy can be presented as a sum of the uniaxial anisotropy energy, the bias field energy, the Zeeman energy in the external field $H_e$, and the domain-wall energy. The minimization procedure results in the following equation for the equilibrium angles:

$$H_a \sin(\theta_j - \psi)\cos(\theta_j - \psi) + H_b \sin(\theta_j - \varphi) + H_e \sin\theta_j = 0, \qquad (2)$$

where $j=1,2$. Equation (2) has two different solutions corresponding to the free-energy minima within some range of the external magnetic field. Within the range, the domain structure may exist. The equilibrium domain-wall displacement $z_0$ can be found from

$$z_0 = (aM/\beta)[H_e\{\cos\theta_1 - \cos\theta_2\} + H_b\{\cos(\theta_1 - \varphi) - \cos(\theta_2 - \varphi)\} \quad (3)$$
$$- (H_a/2)\{\sin^2(\theta_1 - \psi) - \sin^2(\theta_2 - \psi)\}],$$

where $M$ is the saturation magnetization, $a$ is the domain width at zero external magnetic field and in the absence of the bias field, and $\beta$ is the domain-wall pinning parameter. Note that the domain-wall energy is represented by a parabolic potential.[11,12]

The contribution from the domain-walls motion to the transverse susceptibility, $\chi_{dw}$, is found by means of the well-known procedure of the analysis of the domain-wall dynamics in the field of the current,[9,11]

$$\chi_{dw} = \chi_0 (\sin\theta_1 - \sin\theta_2)^2 /(1 - i\omega/\omega_{dw}), \quad (4)$$

where $\chi_0 = aM^2/\beta$ is the static domain-wall susceptibility, $\omega_{dw} = \beta/\alpha$ is the relaxation frequency for the domain-walls motion, and $\alpha$ is the domain-wall mobility proportional to the eddy current losses. The domain-wall mobility is estimated by means of the following expression similar to that obtained in Refs. 9,13:

$$\alpha = 45\sigma dM^2 (\cos\theta_1 - \cos\theta_2)^2 / 2ac^2. \quad (5)$$

The contribution from the magnetization rotation to the susceptibility can be calculated by solving the linearized Landau–Lifshitz equation. In general, the susceptibility is represented by a nondiagonal tensor even after averaging over domains.[9,14] The average transverse rotational susceptibility, $\langle\chi_{rot}\rangle$, can be found from[9]

$$\langle\chi_{rot}\rangle = \langle\chi_2\rangle - 4\pi\langle\chi_3\rangle/(1 + 4\pi\langle\chi_1\rangle), \quad (6)$$

where the averaged susceptibility components $\langle\chi_k\rangle$ ($k=1,2,3$) are given by[14]

$$\begin{aligned}
&\langle\chi_k\rangle = [\chi_k^{(1)} + \chi_k^{(2)}]/2 + [\chi_k^{(1)} - \chi_k^{(2)}](z_0/a), \\
&\chi_1^{(j)} = \gamma M[\omega_1^{(j)} - i\kappa\omega]/\Delta_j, \\
&\chi_2^{(j)} = \gamma M[\omega_2^{(j)} - i\kappa\omega]/\Delta_j, \\
&\chi_3^{(j)} = \gamma M\omega\cos\theta_j/\Delta_j, \\
&\Delta_j = [\omega_1^{(j)} - i\kappa\omega][\omega_2^{(j)} - i\kappa\omega] - \omega^2.
\end{aligned} \quad (7)$$

Here $\gamma$ is the gyromagnetic constant, $\kappa$ is the Gilbert damping parameter, and

$$\begin{aligned}
\omega_1^{(j)} &= \gamma[H_a\cos^2(\theta_j - \psi) + H_e\cos\theta_j + H_b\cos(\theta_j - \varphi)], \\
\omega_2^{(j)} &= \gamma[H_a\cos\{2(\theta_j - \psi)\} + H_e\cos\theta_j + H_b\cos(\theta_j - \varphi)].
\end{aligned} \quad (8)$$

Finally, the transverse permeability of the ribbon can be calculated as $\mu=1+4\pi(\chi_{dw}+<\chi_{rot}>).$[9]

Since we neglect the magnetostatic energy, the nucleation of the domain walls cannot be described in the framework of the model. In this connection, it is assumed further that the domain structure appears to minimize magnetostatic energy, when two different solutions of Eq. (2) exist. Hence, we consider the domain-walls motion and magnetization rotation as nonhysteresis processes. Indeed, in real ribbons, the domain-walls nucleation has a more complex behavior, and as a result, the hysteresis of the GMI response has been observed.[3,5,6]

The calculated GMI profiles are shown in Fig. 2 for two current frequencies $f=\omega/2\pi$ and different values of $H_b$. At low frequencies, the main contribution to the permeability is due to the domain-walls motion. It is well known that in this case the GMI response shows the single-peak behavior. If the bias field $H_b$ equals zero, the GMI profile is symmetric with respect to the external field [dashed line in Fig. 2(a)]. In the presence of $H_b$, the peak value of the GMI response shifts towards positive fields. This is due to the range of external fields, where two-domain configuration appears, also shifts. Note that this prediction of the model is in agreement with the analysis of the permeability spectra found from experimental data.[6] It follows from Fig. 2(a) that at $f=100$ kHz the GMI profile becomes asymmetric and exhibits a steplike increase near peak field (GMI valve) at the presence of the bias field. The asymmetry and the magnitude of the GMI response increase with $H_b$.

At high frequencies, the domain-walls motion is damped by eddy currents, and the magnetization rotation process determines the permeability. The effect of the domain-walls motion on the GMI is essential only in the vicinity of the field, at which the impedance has minimum. At $f=10$ MHz, the GMI profile shows the two-peak behavior. Due to the influence of the bias field, the profile is asymmetric. With the increase of $H_b$, the asymmetry growths and the peak values of $H_e$ shift towards positive fields.

It should be noted that we assume that the direction of the anisotropy in the surface layer may differ from that of the annealing field. This fact is attributed to the influence of the uniaxial anisotropy in the amorphous phase on the crystallization process in the surface layer. As a result, the angle of the unidirectional crystalline field deviates from the ribbon axis and lies within the range of the angles of the uniaxial anisotropy field and the annealing field. Correspondingly, the bias field also deviates from the ribbon axis. Shown in Fig. 3 are the GMI profiles for different current frequencies and bias field angles $\varphi$. If the bias field is along the ribbon axis, the GMI profile shifts towards positive fields and remains symmetric (see dashed lines in Fig. 3). It follows from Fig. 3 that the asymmetry appears, if the bias field deviates from the ribbon axis, and the asymmetry increases with the growth of $\varphi$.

In summary, the model proposed shows that the existence of the bias field and its direction are the main origin for the GMI valve phenomenon at low frequencies and the asymmetric two-peak behavior of the GMI profile at high frequencies. Note that the calculated low-frequency GMI profiles drop more sharply at the right-hand side of the peak in comparison with the experimental data.[6] The disagreement may be related to the spatial distribution of the bias field, which is neglected in the model. However, even the simplified approach developed allows one to explain qualitatively the field and frequency dependences of the GMI profile observed in the experiments.[3–6] The results obtained may be useful to develop GMI-sensor materials with an exchange coupling.

This work was supported by the Korea Science and Engineering Foundation through ReCAMM. N.A.B. acknowledges the support of the Brain Pool Program.

**References**

[1] T. Kitoh, K. Mohri, and T. Uchiyama, IEEE Trans. Magn. **31**, 3137 (1995).

[2] D. P. Makhnovskiy, L. V. Panina, and D. J. Mapps, Appl. Phys. Lett. **77**, 121 (2000).

[3] C. G. Kim, K. J. Jang, H. C. Kim, and S. S. Yoon, J. Appl. Phys. **85**, 5447 (1999).

[4] K. J. Jang, C. G. Kim, S. S. Yoon, and K. H. Shin, IEEE Trans. Magn. **35**, 3889 (1999).

[5] C. G. Kim, C. O. Kim, and S. S. Yoon, J. Magn. Magn. Mater. **249**, 293 (2002).

[6] Y. W. Rheem, L. Jin, S. S. Yoon, C. G. Kim, and C. O. Kim, IEEE Trans. Magn. **39**, 3100 (2003).

[7] C. G. Kim, K. J. Jang, D. Y. Kim, and S. S. Yoon, Appl. Phys. Lett. **75**, 2114 (1999).

[8] L. Kraus, *Abstracts of the International Conference on Magnetism* (Roma, Italy, 2003) p. 712.

[9] L. V. Panina, K. Mohri, T. Uchiyama, M. Noda, and K. Bushida, IEEE Trans. Magn. **31**, 1249 (1995).

[10] L. Kraus, J. Magn. Magn. Mater. **195**, 764 (1999).

[11] F. L. A. Machado and S. M. Rezende, J. Appl. Phys. **79**, 6558 (1996).

[12] D. Atkinson and P. T. Squire, J. Appl. Phys. **83**, 6569 (1998).

[13] D. X. Chen, J. L. Munoz, A. Hernando, and M. Vazquez, Phys. Rev. B **57**, 10699 (1998).

[14] D. P. Makhnovskiy, L. V. Panina, and D. J. Mapps, Phys. Rev. B **63**, 144424 (2001).

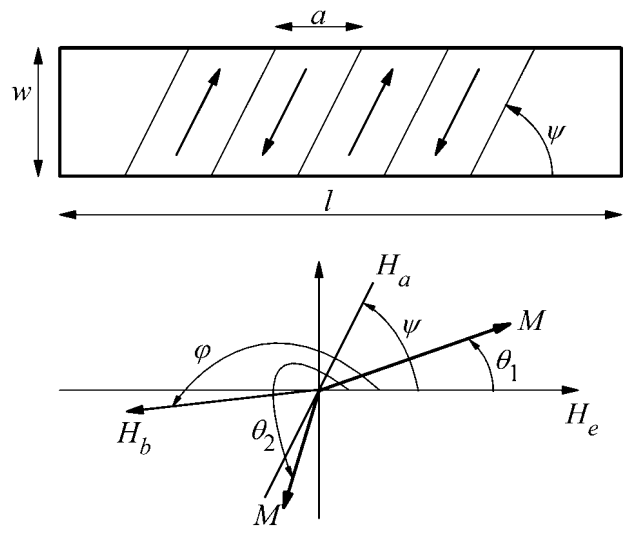

FIG. 1. A sketch of domain structure and coordinate system used for analysis.

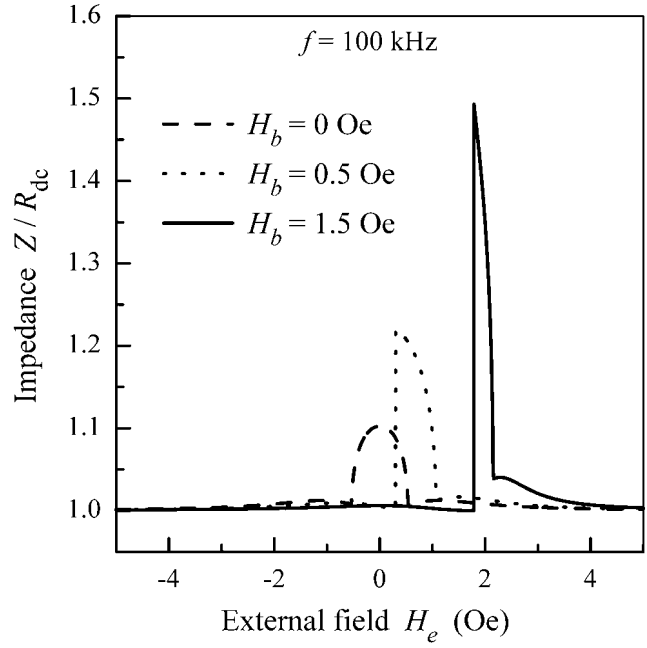

(a)

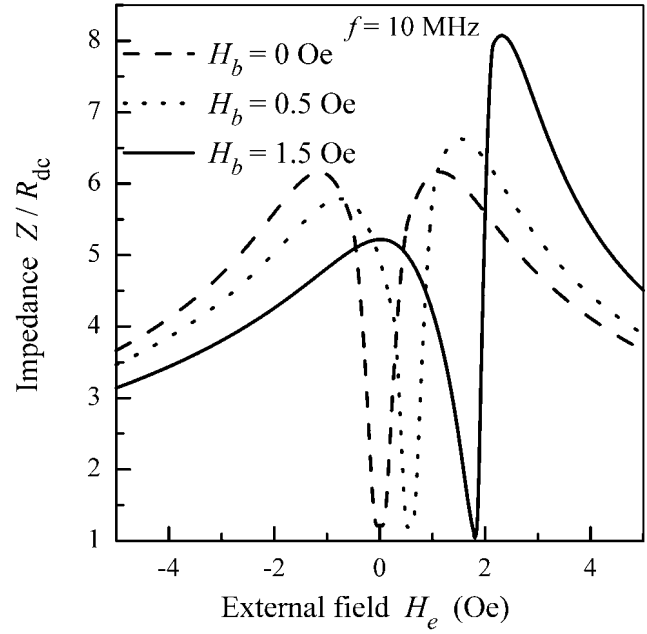

(b)

FIG. 2. GMI profile for different $H_b$ at current frequency $f=100$ kHz (a) and $f=10$ MHz (b). Parameters used for calculations are $d=20\,\mu m$, $a=5\,\mu m$, $\sigma=10^{16}\,s^{-1}$, $M=600$ G, $H_a=1$ Oe, $\beta/MH_a a=0.5$, $\kappa=0.1$, $\psi=0.35\pi$, $\varphi=1.05\pi$.

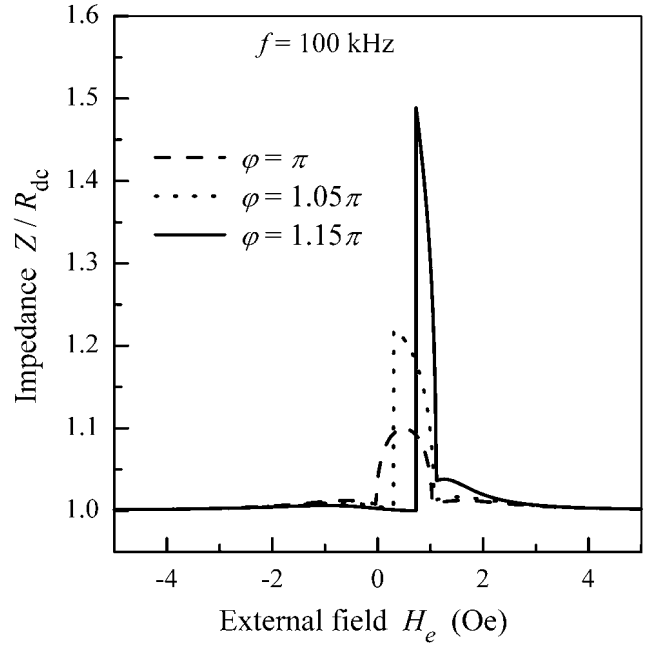

(a)

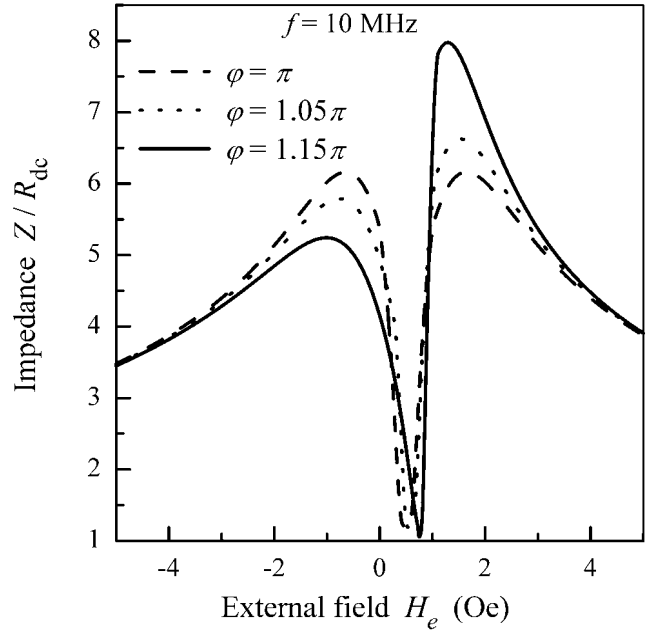

(b)

FIG. 3. GMI profile for different $\varphi$ at current frequency $f=100$ kHz (a) and $f=10$ MHz (b). Parameters used for calculations are $d=20$ μm, $a=5$ μm, $\sigma=10^{16}$ s$^{-1}$, $M=600$ G, $H_a=1$ Oe, $H_b=0.5$ Oe, $\beta/MH_a a=0.5$, $\kappa=0.1$, $\psi=0.35\pi$.